\documentclass[sigconf]{acmart}
\usepackage{subfiles}
\usepackage{xcolor}
\usepackage{multirow}
\usepackage{enumitem}
\usepackage{listings} 
\usepackage{hyperref}
\usepackage{graphicx}
\usepackage{pifont}

\setcopyright{none}
\setcopyright{none}
\renewcommand\footnotetextcopyrightpermission[1]{} 

\AtBeginDocument{%
  \providecommand\BibTeX{{%
    \normalfont B\kern-0.5em{\scshape i\kern-0.25em b}\kern-0.8em\TeX}}}

\usepackage{url}
\begin{document}

\title{S2AMP: A High-Coverage Dataset of Scholarly Mentorship Inferred from Publications}


\author{Shaurya Rohatgi}
\email{szr207@psu.edu}
\affiliation{%
  \institution{Allen Institute for AI and Pennsylvania State University}
  \city{Seattle, WA}
  \country{US}
}

\author{Doug Downey}
\email{dougd@allenai.org}
\affiliation{%
  \institution{Allen Institute for AI}
  \city{Seattle, WA}
  \country{US}
}

\author{Daniel King}
\email{daking@hey.com}
\affiliation{
    \city{Seattle, WA}
    \country{US}
}
\author{Sergey Feldman}
\email{sergey@allenai.org}
\affiliation{%
  \institution{Allen Institute for AI}
  \city{Seattle, WA}
  \country{US}
}


\renewcommand{\shortauthors}{Rohatgi, et al.}

\begin{abstract}
Mentorship is a critical component of academia, but is not as visible as publications, citations, grants, and awards. Despite the importance of studying the quality and impact of mentorship, there are few large representative mentorship datasets available. We contribute two datasets to the study of mentorship. The first has over 300,000 ground truth academic mentor-mentee pairs obtained from multiple diverse, manually-curated sources, and linked to the Semantic Scholar (S2) knowledge graph. We use this dataset to train an accurate classifier for predicting mentorship relations from bibliographic features, achieving a held-out area under the ROC curve of 0.96. Our second dataset is formed by applying the classifier to the complete co-authorship graph of S2. The result is an inferred graph with 137 million weighted mentorship edges among 24 million nodes. We release this first-of-its-kind dataset to the community to help accelerate the study of scholarly mentorship: \url{https://github.com/allenai/S2AMP-data}
\end{abstract}




\maketitle
\pagestyle{plain}

\section{Introduction}
    


Mentorship is a crucial part of the scholarly enterprise. A mentor who has gained experience working in a field over time guides a mentee who is new to the area, and this phase can have long-term effects throughout a mentee's career \cite{Yifang1073, Linard2018IntellectualSI, Sinatra2016,Li2019EarlyCW}.  Evidence of effective mentorship can be an important factor in promotion and tenure decisions or in a student's choice of advisor.  However, records of mentorship are not available at scale the way standard bibliographic measures like h-index, citation count and number of published papers are. To remedy this we introduce two related datasets for studying and inferring publication-evidenced mentorship in science at scale, which we collectively refer to as Semantic Scholar Analysis of MentorshiP (S2AMP).

S2AMP is aimed at identifying explicit mentorship relations, such as those between Ph.D. advisors and their students or senior research managers and their less senior coworkers.  We start by aggregating Web resources containing ground truth mentorship relations of this type.  These form our first dataset.  Our second dataset is then obtained by applying a mentorship classifier trained on the first dataset to infer mentorship relations across a complete bibliographic knowledge graph.  We find that accurate inference of mentorship is possible, and that the inferred dataset greatly exceeds the coverage of previous mentorship datasets. The largest similar resource is the crowd-sourced Academic Family Tree (AFT) \cite{Ke2021} linked to the Microsoft Academic Graph (MAG), which is approximately 180 times smaller than our inferred dataset in terms of the number of mentor-mentee pairs (see Table \ref{tab:teaser-table}).  Other previous work, like SHIFU \cite{Wang1145, Liu2020, Lee2020}  built mentorship models using publication information provided by MAG \cite{wang2020microsoft}. This model was then used to infer mentorship over a part of the MAG (one million authors). However, this inferred data was not open sourced by the authors and only applied to a small portion of all candidate author pairs. Other work which models mentorship has focused on data for a specific field of study like Physics \cite{Heinisch2018TheNG} or Mathematics and AI \cite{Rao2010}. 

\begin{table}[t]
\begin{small}
\caption{Comparison of S2AMP datasets with Academic Family Tree (AFT) dataset linked subset.}
\label{tab:teaser-table}
\begin{tabular}{c|ccc}
& \textbf{\begin{tabular}[c]{@{}c@{}}AFT\\Linked\end{tabular}} & \textbf{\begin{tabular}[c]{@{}c@{}}S2AMP\\ Gold\end{tabular}} & \textbf{\begin{tabular}[c]{@{}c@{}}S2AMP\\ Inferred\end{tabular}} \\ 
\midrule
\textbf{\# Mentor-Mentee Pairs} & 743,176 & 330,576 & \textbf{137M} \\
\textbf{\# of Scholars} & 738,989 & 356,541 & \textbf{24M} \\ \hline
\textbf{\begin{tabular}[c]{@{}c@{}}author metadata\\  source\end{tabular}} & \begin{tabular}[c]{@{}c@{}}MAG\\ (Deprecated)\end{tabular}        & \begin{tabular}[c]{@{}c@{}}Semantic \\ Scholar\end{tabular}         & \begin{tabular}[c]{@{}c@{}}Semantic \\ Scholar\end{tabular}         \\ \hline
\textbf{Co-Publication Info}                                                & no                                                                & \textbf{yes}                                                        & \textbf{yes}                                                       
\end{tabular}
\end{small}
\end{table}

Our new S2AMP resource includes the following contributions:
\begin{enumerate}[leftmargin=*]
    \item {\em S2AMP Gold} - A ground truth dataset drawn from multiple online sources containing known mentor-mentee pairs, and links to rich bibliographic records of the individuals in Semantic Scholar (S2)  \cite{Ammar2018ConstructionOT}.\footnote{Previous mentorship resources linked to MAG for relevant metadata instead, but it has been deprecated as of December, 2021.}
    \item A classifier that infers mentorship relations from bibliographic data based on the bibliographic records of the pair (e.g. inferred seniority, co-publication, and author order), and graph-based features derived from an initial estimated mentorship graph (e.g., estimated number of mentees).  The final classifier is efficient, and has an area under the ROC curve of 0.96. The area under the precision-recall curve is 0.74, and a manual analysis (detailed in Section \ref{section:error_analysis}) of 100 false positives revealed that 92\% of them were in fact true positives that were simply absent from the gold data.
    \item {\em S2AMP Inferred} - A mentor-mentee dataset of 137 million inferred mentorship relations obtained by applying the trained classifier to the entire S2 knowledge graph.
\end{enumerate}

The rest of the paper proceeds as follows.  In Section \ref{data}, we detail how we collect ground truth mentor-mentee relations and look up the pairs of authors in S2. In Section \ref{methods} we describe how we create negative examples for each true mentor-mentee pair to train our classifier.  We also show how a novel two-stage model, that first classifies pairs individually and then re-classifies them using graph features derived from the first stage, can improve accuracy (Section \ref{modeling}). Finally, in Section \ref{graph} we discuss the creation of our inferred large-scale mentorship graph, and provide preliminary analyses using this graph to illustrate its utility for studying mentorship.


\section{Data Collection}\label{data}

The first step in our data collection is to acquire ground-truth mentor-mentee pairs from online sources.  Specifically, we build custom crawlers for Open ProQuest\footnote{\url{https://www.pqdtopen.proquest.com/}}, The Mathematics Genealogy Project\footnote{\url{https://www.genealogy.math.ndsu.nodak.edu/}} and Handle.net\footnote{\url{https://handle.net/}} pages associated with major Universities around the world. We obtain approximately 280,000 pairs from these sources. Lastly, we use the Academic Family Tree\footnote{\url{https://academictree.org/}} dataset. This dataset has 1.5 million mentor-mentee pairs populated by end-users (of which 743,176 are linked to MAG in \citet{Ke2021}).

\paragraph{Linking to Semantic Scholar}
Our mentorship classifier relies on the bibliographic records of the mentors and mentees.  To obtain these records, we link our gold mentor-mentee pairs to author nodes in an existing bibliographic knowledge base, Semantic Scholar (S2).  We first links the mentees and then the mentors, as described below.

Given a mentor-mentee pair, and one or more author records from S2 that may correspond to the mentee, we find the mentor by searching through each of the candidate author's co-authors for a textual match for the mentor.  We retain the mentee id for which there is a matching mentor name. In case of multiple matches, we keep the pairs with the most co-authored papers.

Using the above procedure, we successfully match over 300,000 mentor and mentee pairs to S2's authorship graph. For each matched author, we obtain their co-authors and complete publication history using S2.\footnote{As of March 2022, S2 had 204 million scholarly articles and 76 million authors linked with their article records.}  These linked ground-truth mentor-mentee pairs and their bibliographic data comprise {\em S2AMP-Gold}.





\paragraph{Linking Errors} 
The linking procedure suffers from occasional errors. Picking between ambiguous name pairs by using the "most co-published papers" heuristic is sometimes wrong, and sometimes the linking may be correct but the author profiles themselves incorrect due to automated author disambiguation errors. 

\section{Methodology}\label{methods}




The linked pairs described in the previous section serve as positive mentor-mentee pairs for model training. Next, we describe how we obtain negative examples and the features used to train our model.

\subsection{Actual and Dense Co-Publication Period}
We define the {\em co-publication period} as the time interval (in years) between the first co-authored article and the latest co-authored article. This co-publication period may be longer than the period of explicit mentorship.  For example, cases when a Ph.D. student graduated, became a professor, then later collaborated again with their Ph.D. advisor. For these types of cases, it is unclear when to call a mentorship complete from the publication history alone. As such, we also define a \emph{dense} co-publication period as the shortest period during the authors published P\% (P>60) of their co-publications, and use this as a proxy for the most important collaboration period between a pair of authors. When we derive the features for each pair, we compute two versions of each feature, one for the co-publication period and another for the dense co-publication period. 

\subsection{Candidate Mentors}
\label{candidate_mentors}
To construct negative examples, we obtain the co-authors of a mentee who are plausibly mentors but not marked as such in the ground truth data. These may include, for example, other more senior Ph.D. students, post-doctoral students, or other co-author professors. We call them {\em candidate mentors}, and they serve as negative examples for the classification model. 

Using the bibliographic data, we select as candidate mentors all the co-authors of a mentee where (a) the co-author has more publications than the mentee before the date of their first co-authored article, (b) the co-author's first publication came earlier than the mentee's, and (c) the co-author and the mentee have written at least $k$ scholarly articles together. In our experiments we set $k = 2$, after finding that $k=1$ yielded significantly lower precision and a computationally prohibitive number of candidate mentors during the inference stage. Requiring two co-publications can be helpful for capturing short-term mentorships for pre-doctoral mentees in some high-publication fields, in addition to longer-term mentorships with Ph.D. students in many other fields that publish before graduation.

\subsection{Pairwise Features: First Stage}

Using author and publication data from S2, we extract various types of features for each positive and negative mentor-mentee pair:

\begin{itemize}[leftmargin=*]
\item{Co-publication} - Features based on the co-publications of the mentor and the mentee, as well as total individual publications of the mentor and mentee during the co-publication period. These features capture our expectations that, for example, the mentor will have a significantly higher number of publications than the mentee for a given period, and that the mentee will have fewer years of scholarly activity than their mentor. 
\item{Co-authors} - The count of co-authors for each mentor and the mentee. We expect mentors to typically have more co-authors or collaborators than mentees for the period of co-publication.
\item{Author list position} - We use the authorship position of the mentee and the mentor in the publications which they co-author as well as the ones they don't for the period of co-publication. This can be a helpful feature in fields where mentors tend to occur at the end of the authorship list.\footnote{This is not true in all fields. We attempted to account for this by including field of study as an explicit feature, but it had no effect on model performance.}
\end{itemize}
We curate a total of 48 features, which are detailed in the Appendix in Table \ref{tab:feature-desc}. Using these features we train a classifier (described in the next section) to predict if a given pair of authors are a positive mentor-mentee pair or not. 

\subsection{Graph-Based Features: Second Stage}
The model trained on features from the previous section predicts a mentor-mentee relationships with no context about surrounding colleagues, and this is a limitation. In practice, mentors tend to mentor more than one mentee over the course of their career, and a mentee who already has a high-scoring mentor is less likely to have others with as high a score. As such, we can improve our model's first-stage predictions by taking into account other nearby predictions for each candidate mentor and mentee. 

Formally, we define a graph where each candidate mentor and mentee correspond to nodes, and weighted directed edges point from each mentor to each mentee (where the edge weight is the score produced by the first-stage model). We extract additional features for each node in the graph. For example, for each node we sum the \emph{incoming} edge weights to get the total ``menteeship'' received, and sum the \emph{outgoing} edges to obtain the total mentorship given. We extract a total of 20 features using the first-stage mentorship graph, including the maximum mentorship given by an author, ratio of sum of in-edge weights to out-edge weights, difference between the sum of in-edge weights and out-edge weights, and others, all of which are described in the Appendix in Table \ref{tab:feature-desc}. These new features are concatenated with the first-stage features and a second-stage model is trained with the same methodology as the first-stage model.

\section{Modeling}\label{modeling}


Filtering our gold mentor-mentee pairs from our linking pipeline to just those pairs that have co-authored at least $k=2$ papers leaves a subset of 219,331 positive pairs, to which we add the generated 1.6 million negative mentor-mentee pairs. We then split the data into train, validation, and test (3:1:1 split), ensuring that mentees present in one split are not present in another. 

Using the extracted features we train a binary classification model to differentiate true mentor-mentee pairs from the false ones. As this model is eventually used for inference at scale for approximately 137 million pairs, we chose the efficient LightGBM \cite{ke2017lightgbm} implementation of gradient boosted decision tree ensembles. Evaluating on the validation set, we use the \texttt{hyperopt} library \cite{bergstra2013making} to search for the best configuration of nine hyperparameters.\footnote{See Figure \ref{fig:hyperopt} in the Appendix for complete details.} We use the probabilistic output from the trained LightGBM classifier as the edge weight in the mentorship graph. To report final performance on the test set, we use two metrics: area under the ROC curve (AUROC) and area under the precision recall curve (AUPRC). Results for the first and second stages are reported in Table \ref{tab:evaluation}.

We also study what features are important for predicting a mentorship relationship using SHAP importance \cite{NIPS2017_7062} values. We observe that the number of publications before the co-publication period for both mentors and mentees, number of coauthors before the co-publication period for both mentors and mentees, and average of authorship position of the mentor are the top-3 features for the first-stage model. For the second-stage, the two most important features include node out-edge weight average of the mentor (average mentorship given) and sum of in-edge weights for the mentee (total mentorship received). See Figures \ref{fig:shap1} and \ref{fig:shap2} in the Appendix for a complete visualization of the SHAP values.

\begin{table}[t]
\caption{\label{tab:evaluation} Performance of the classifier for both stages, and with features categories added iteratively. More complex features and the second stage classifier considerably improve performance.}
\vspace{-15pt}
\begin{tabular}{llll}
                                           & \textbf{Features}                & \textbf{AUPRC}  & \textbf{AUROCC} \\ 
\midrule
\multicolumn{1}{c|}{\textbf{}}             & \multicolumn{1}{l|}{publication} & 0.5795          & 0.9171         \\
\multicolumn{1}{l|}{\textbf{First Stage}}  & \multicolumn{1}{l|}{+ coauthor}  & 0.6778          & 0.9444          \\
\multicolumn{1}{l|}{}                      & \multicolumn{1}{l|}{+ position}  & 0.7001          & 0.9519         \\ \cline{1-1}
\multicolumn{1}{l|}{\textbf{Second Stage}} & \multicolumn{1}{l|}{+ graph}     & \textbf{0.7368} & \textbf{0.9563} \\
\end{tabular}

\end{table}

\section{Mentorship Graph}\label{graph}

The complete co-publication graph of Semantic Scholar contains around 2 billion edges. We use the criteria from Section \ref{candidate_mentors} to select candidate mentors for each mentee from these co-publication edges, after which we are left with 137 million pairs of mentees and their candidate mentors. These pairs are run through our feature extraction pipeline and the trained two-stage model to obtain a set of predicted scores, which serve as weights on the inferred mentorship edges. We can then calculate the following mentorship metrics for each author node:

\begin{itemize}[leftmargin=*]
\item {\em Menteeship sum} - sum of all the incoming edges of a node (mentorship received).
\item {\em Menteeship mean} - mean of all the incoming edges.
\item {\em Mentorship sum} - sum of all the outgoing edges of a node (mentorship given).
\item {\em Mentorship mean} - sum of all the outgoing edges.
\end{itemize}

\subsection{Graph Analysis via Fixed Effects Modeling}
We hypothesize that the derived mentorship scores from the preceding section not only meaningfully quantify mentorship given and received, but can also partially explain success in academia more broadly, as measured by h-index. We thus model the relationship between h-index (the dependent variable) and total publication count, total citation count, field of study, and the four derived mentorship scores (independent variables)\footnote{We release the bibliographic metrics and the field of study along with our dataset. These values are obtained from S2.} with a negative binomial general linear fixed effects model\footnote{These GLMs are commonly used to model count-based data where the variance is greater than the mean, which is the case for h-index across authors in our inferred dataset.}, the conditional mean model of which is
$E[\text{h-index}|\mathbf{x}] = \text{exp}\left(w_0 + \sum_i w_i\cdot x_i \right),$
where $\mathbf{x}$ is the vector of covariates and $w_i$ is the learned coefficient for the $i$th covariate. We represent the field of study covariate using one-hot encoding, while the continuous variables are each binned into five equally-sized quintiles. We fit the model using the entire inferred graph (mentors and mentees both) after removing outliers, and the learned coefficients are in Table \ref{tab:fixed_effects}. Important to note is that, due to the exponential term in the regression function above, the coefficients are interpreted as having a \emph{multiplicative} effect instead of an additive one as in linear regression. For example, the coefficient for the 5th quintile of menteeship sum is 0.14, which we interpret as having the effect of multiplying the h-index by exp(0.14) = 1.14.  The first quintile of each covariate has 0 coefficient by construction. We exclude coefficients for field of study to save space, but Table \ref{tab:feature-desc} in the Appendix has complete coefficients information.

The coefficients for citation count are the largest, followed by paper count, as expected since the h-index is constructed from these two variables. Menteeship sum is the next largest, statistically significant variable: as the menteeship sum increases, its coefficient for predicting h-index also increases. Interestingly, both menteeship mean and mentorship sum are negative, with the exception of the 5th quintile of mentorship sum. Mentorship mean is a small, constant, but statistically positive effect. Menteeship sum and mentorship mean are also the two most important covariates for predictive power according to SHAP. We note that causal conclusions should not be made from this analysis.  

\begin{table}[t]
\caption{Coefficients for the negative binomial GLM with outcome variable h-index. A * indicates the entry is significantly different from 0 with p-value \textless 0.0001.  All  mentorship features are significantly predictive of h-index, even after accounting for citation, paper count and field of study.}
\vspace{-5pt}
\label{tab:fixed_effects}
\begin{tabular}{lllllll}
         &  &   & \multicolumn{2}{l}{\textbf{Menteeship}}      & \multicolumn{2}{l}{\textbf{Mentorship}} \\ 
\midrule
\textbf{Quintile} & \begin{tabular}[c]{@{}l@{}}\textbf{Citation}\\ \textbf{Count}\end{tabular} & \multicolumn{1}{l|}{\begin{tabular}[c]{@{}l@{}}\textbf{Paper}\\ \textbf{Count}\end{tabular}} & \textbf{Sum}   & \multicolumn{1}{l|}{\textbf{Mean}}   & \textbf{Sum}            & \textbf{Mean}          \\ 
\midrule
1st & 0     & \multicolumn{1}{l|}{0}     & 0     & \multicolumn{1}{l|}{0}      & 0      & 0 \\
2nd & 1.83* & \multicolumn{1}{l|}{0.21*} & 0.06* & \multicolumn{1}{l|}{-0.02*} & -0.01  & 0.03* \\
3rd & 2.46* & \multicolumn{1}{l|}{0.33*} & 0.08* & \multicolumn{1}{l|}{-0.03*} & -0.01* & 0.05* \\
4th & 2.92* & \multicolumn{1}{l|}{0.44*} & 0.11* & \multicolumn{1}{l|}{-0.04*} & -0.01* & 0.05* \\
5th & 3.51* & \multicolumn{1}{l|}{0.7*}  & 0.14* & \multicolumn{1}{l|}{-0.02*} &  0.1*  & 0.03*  \\ 
\end{tabular}

\end{table}

\subsection{Mentor and Mentee Discovery}
Mentorship scores can help us identify expert scholars and their descendants. For example, our model predicts a high mentorship sum (272.44) for Leo Paquette\footnote{\url{https://en.wikipedia.org/wiki/Leo_Paquette}}, an American organic chemist who guided approximately 150 graduate students to their Ph.D. degree. To highlight the coverage of our model, S2AMP has 279 mentees (above a mentorship score of 0.5) for Dr. Paquette while Academic Tree reports only 39.\footnote{\url{https://academictree.org/chemistry/peopleinfo.php?pid=55614}} We note again that these are the mentees detectable by our algorithm, and it is likely Leo Paquette mentored others with whom he published zero or one paper. More examples of expert discovery are discussed in Appendix. 

\subsection{Error Analysis}
\label{section:error_analysis}
To assess the quality of our model, we manually examined 200 misclassified sample pairs from the validation data (100 false positives and 100 false negatives). We found via manual web search and author homepage review that 92\% of the false positives were true mentor-mentee pairs. The other 8\% were cases where the candidate mentor was very senior to the mentee, but did not appear to be a mentor. Looking at the false negatives, we found that the model gives low probabilities to pairs with mentors who have not mentored many other mentees, which is common for mentors who have just started their careers.

\section{Conclusion}
We introduce S2AMP, a pair of novel datasets for studying publication evidenced mentorship. S2AMP Gold is used to train a model which can generalize mentorship prediction to unseen pairs. We used this model to generate S2AMP Inferred, a graph with 137 million edges and 24 million nodes. The Gold and Inferred datasets are linked to S2, which is regularly updated. We hope that S2AMP helps further new studies and findings in understanding the role of mentorship in academia and beyond.

\paragraph{Limitations}
As discussed in the introduction, S2AMP targets {\em explicit} mentorship relationships, such as Ph.D advising or management in industry.  However, this is not the only type of mentorship that occurs in scholarly work.  Mentorship can be informal and not explicitly documented, e.g. when more senior lab mates mentor junior ones. Our ground truth data does not cover this informal mentorship.  

Similar to the previous work \cite{wang2010mining}, our model can't detect mentorship unless there is co-publication, and we require a minimum of two co-published papers. This naturally excludes early academic career mentorship detection in many fields where publishing during a Ph.D. is not as common, e.g. humanities and economics.  Both S2AMP Gold and S2AMP Inferred exclude any form of mentorship which has no publication trace.


\paragraph{Future Work}
We believe that our inferred mentorship graph can help answer many science-of-science \cite{Yifang1073, Schwartz2021} questions including: 
\begin{itemize}[leftmargin=*]
    \item Do doctoral students whose advisors have superior research output and/or are better positioned in the co-publication network of their respective scientific communities have a higher probability to become advisors themselves? \cite{Heinisch2017TheNG}
    \item How does mentorship differ across different fields of study? Are cross-disciplinary mentorships correlated with success?
    \item Can we identify significant ``intermediate actors'' that play a role in the mentorship graph, e.g. post-doctoral students?
\end{itemize}

One important question is to what degree our model can discover informal mentorship relationships in addition to formal ones like Ph.D. advisors and their students. The error analysis in Section \ref{section:error_analysis} showed promising results for discovering formal relationships.  We speculate that informal relationships will be reflected by somewhat elevated scores from our model, and measuring this is an item of future work.

\section{Acknowledgements}
We are immensely grateful to James Thorson, Jevin West, Dashun Wang, and Daniel S. Weld for their insightful discussions, and JJ Yang for their relentless support.  This work was supported in part by NSF Convergence Accelerator Grant 2132318.

\bibliographystyle{ACM-Reference-Format}
\bibliography{references}

\clearpage

\appendix

\section{Mentor and Mentee Discovery}
\label{discovery}
Here are some examples for scholars with high mentorship or menteeship sum score:
\begin{itemize}
    
    \item \textbf{Author Name}: Yuen Kwok-yung \\
    \textbf{h-index}: 135 \\
    \textbf{Mentorship sum}: 204.6  \\
    \textbf{Field of study}: Medicine\\
    "He led a team identifying the SARS coronavirus that caused the SARS pandemic of 2003–4, and traced its genetic origins to wild bats. During the ongoing COVID-19 pandemic, he has acted as expert adviser to the Hong Kong government."\footnote{\url{https://en.wikipedia.org/wiki/Yuen_Kwok-yung}}
    
    \item \textbf{Author Name}: Michael J. Black  \\
    \textbf{h-index}: 96 \\
    \textbf{Menteeship sum}: 4.8  \\
    \textbf{Field of study}: Computer Science\\
    "[T]he only researcher in the field to have won all three major test-of-time prizes in computer vision: the 2010 Koenderink Prize at the European Conference on Computer Vision (ECCV), the 2013 Helmholtz Prize at the International Conference on Computer Vision (ICCV), and the 2020 Longuet-Higgins Prize at the IEEE Conference on Computer Vision and Pattern Recognition (CVPR)."\footnote{\url{https://en.wikipedia.org/wiki/Michael_J._Black}}
    
    \item \textbf{Author Name}: Kaushik Roy \\
    \textbf{h-index}: 96 \\
    \textbf{Mentorship sum}: 127.2  \\
    \textbf{Field of study}: Electrical Engineering\\
    "He has supervised 91 Ph.D. dissertations and co-authored two books on Low Power CMOS VLSI Design (John Wiley \& McGraw Hill)."\footnote{\url{https://en.wikipedia.org/wiki/Kaushik_Roy}}
    
    \item \textbf{Author Name}:  Peter Nijkamp \\
    \textbf{h-index}: 96 \\
    \textbf{Mentorship sum}: 169.8  \\
    \textbf{Field of study}: Economics\\
    "He is ranked among the top 100 economists in the world according to IDEAS/RePEc, and is by far the most prolific economist."\footnote{\url{https://en.wikipedia.org/wiki/Peter_Nijkamp}}
\end{itemize}

\begin{figure}[h]
  \centering
  \includegraphics[width=\linewidth]{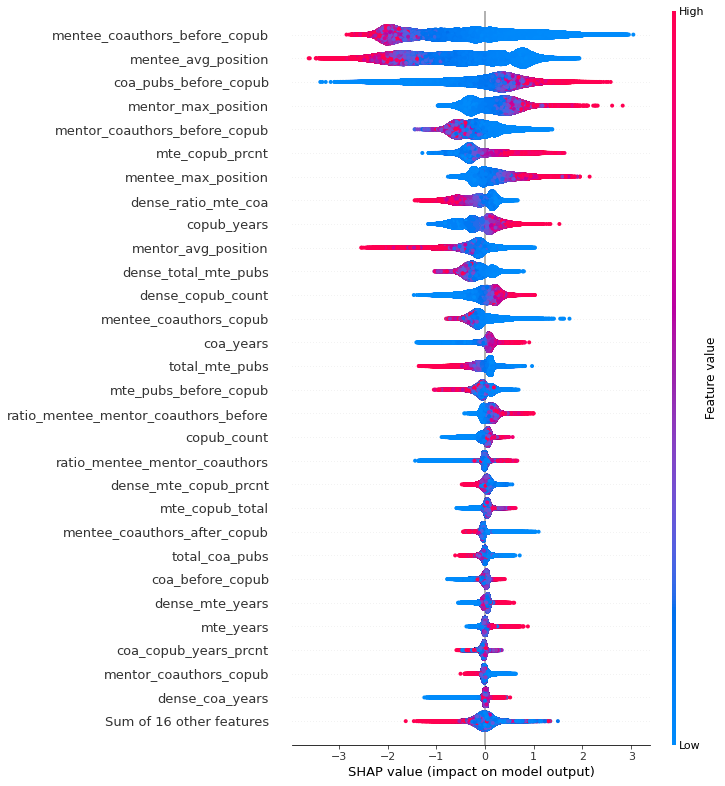}
  \caption{\label{fig:shap1}SHAP plot for first-stage pairwise features. The feature with the highest value is 
    \textit{mentee\_coauthor\_before\_copub} interpreted as: higher value of this feature decreases the probability of mentorship.}
\end{figure}
\begin{figure}[h]
  \centering
  \includegraphics[width=\linewidth]{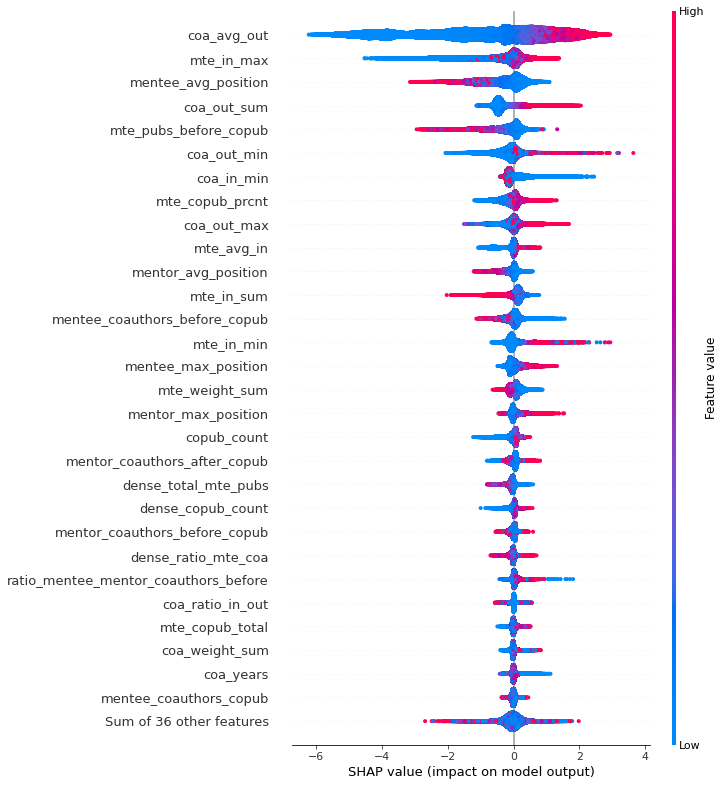}
  \caption{\label{fig:shap2}SHAP plot for second-stage graph features. The feature with the highest value is \textit{coa\_avg\_out} (mean mentorship score) interpreted as: higher value of this feature increases the probability of mentorship.}
\end{figure}

\begin{table*}[h]
\begin{small}
\caption{Description for features extracted from publication information retrieved from Semantic Scholar.}
\label{tab:feature-desc}
\begin{tabular}{ccc}
\textbf{Category}                                           & \textbf{Name}                                     & \textbf{Description}                                                         \\ \hline
\multicolumn{1}{c|}{\multirow{27}{*}{\textbf{publication}}} & \textit{copub\_count}                             & total papers published together                                              \\
\multicolumn{1}{c|}{}                                       & \textit{total\_mte\_pubs}                         & total publications of the mentee till copub end date                         \\
\multicolumn{1}{c|}{}                                       & \textit{total\_coa\_pubs}                         & total publications of the coauthor till copub end date                       \\
\multicolumn{1}{c|}{}                                       & \textit{mte\_copub\_total}                        & \# of papers published by mentee in copub period                             \\
\multicolumn{1}{c|}{}                                       & \textit{coa\_copub\_total}                        & \# of papers published by coauthor in copub period                           \\
\multicolumn{1}{c|}{}                                       & \textit{mte\_copub\_prcnt}                        & ratio of copub\_count to mte\_copub\_total                                   \\
\multicolumn{1}{c|}{}                                       & \textit{coa\_copub\_prcnt}                        & ratio of copub\_count to coa\_copub\_total                                   \\
\multicolumn{1}{c|}{}                                       & \textit{ratio\_mte\_coa}                          & ratio of total\_mte\_pubs to total\_coa\_pubs                                \\
\multicolumn{1}{c|}{}                                       & \textit{copub\_years}                             & \# of years of collaboration                                                 \\
\multicolumn{1}{c|}{}                                       & \textit{mte\_years}                               & mentee publication years till copub end date                                 \\
\multicolumn{1}{c|}{}                                       & \textit{coa\_years}                               & coauthor publication years till copub end date                               \\
\multicolumn{1}{c|}{}                                       & \textit{mte\_copub\_years\_prcnt}                 & ratio of copub\_years to mte\_years                                          \\
\multicolumn{1}{c|}{}                                       & \textit{coa\_copub\_years\_prcnt}                 & ratio of copub\_years to coa\_years                                          \\
\multicolumn{1}{c|}{}                                       & \textit{dense\_mte\_copub\_total}                 & \# of papers published by mentee in dense copub period                       \\
\multicolumn{1}{c|}{}                                       & \textit{dense\_coa\_copub\_total}                 & \# of papers published by coauthor in dense copub period                     \\
\multicolumn{1}{c|}{}                                       & \textit{dense\_total\_coa\_pubs}                  & total publications of the coauthor till dense copub end date                 \\
\multicolumn{1}{c|}{}                                       & \textit{dense\_total\_mte\_pubs}                  & total publications of the mentee till dense copub end date                   \\
\multicolumn{1}{c|}{}                                       & \textit{dense\_copub\_count}                      & total papers published together during the dense copub period                \\
\multicolumn{1}{c|}{}                                       & \textit{dense\_mte\_copub\_prcnt}                 & ratio of dense copub\_count to mte\_copub\_total                             \\
\multicolumn{1}{c|}{}                                       & \textit{dense\_coa\_copub\_prcnt}                 & ratio of dense copub\_count to coa\_copub\_total                             \\
\multicolumn{1}{c|}{}                                       & \textit{dense\_ratio\_mte\_coa}                   & ratio of dense\_total\_mte\_pubs to dense\_total\_coa\_pubs                  \\
\multicolumn{1}{c|}{}                                       & \textit{dense\_mte\_years}                        & mentee publication years till dense copub end date                           \\
\multicolumn{1}{c|}{}                                       & \textit{dense\_coa\_years}                        & coauthor publication years till dense copub end date                         \\
\multicolumn{1}{c|}{}                                       & \textit{dense\_mte\_copub\_years\_prcnt}          & ratio of dense copub\_years to mte\_years                                    \\
\multicolumn{1}{c|}{}                                       & \textit{dense\_coa\_copub\_years\_prcnt}          & ratio of dense copub\_years to coa\_years                                    \\
\multicolumn{1}{c|}{}                                       & \textit{coa\_pubs\_before\_copub}                 & coauthor publication count before co-publication period                      \\
\multicolumn{1}{c|}{}                                       & \textit{mte\_pubs\_before\_copub}                 & mentee publication count before co-publication period                        \\ \hline
\multicolumn{1}{c|}{\multirow{8}{*}{\textbf{co-author}}}    & \textit{mentee\_coauthors\_before\_copub}         & number of coauthors of mentee before copublication period                    \\
\multicolumn{1}{c|}{}                                       & \textit{mentor\_coauthors\_before\_copub}         & number of coauthors of mentor before copublication period                    \\
\multicolumn{1}{c|}{}                                       & \textit{mentee\_coauthors\_after\_copub}          & number of coauthors of mentee at the end of copublication period             \\
\multicolumn{1}{c|}{}                                       & \textit{mentor\_coauthors\_after\_copub}          & number of coauthors of mentor at the end of copublication period             \\
\multicolumn{1}{c|}{}                                       & \textit{mentor\_coauthors\_copub}                 & number of coauthors of mentee during copublication period                    \\
\multicolumn{1}{c|}{}                                       & \textit{ratio\_mentee\_mentor\_coauthors}         & mentee\_coauthors\_copub divided by mentor\_coauthors\_copub                 \\
\multicolumn{1}{c|}{}                                       & \textit{ratio\_mentee\_mentor\_coauthors\_before} & mentee\_coauthors\_before\_copub divided by mentor\_coauthors\_before\_copub \\
\multicolumn{1}{c|}{}                                       & \textit{ratio\_mentee\_mentor\_coauthors\_after}  & mentee\_coauthors\_after\_copub divided by mentor\_coauthors\_after\_copub   \\ \hline
\multicolumn{1}{c|}{\multirow{6}{*}{\textbf{position}}}     & \textit{mentee\_min\_position}                    & min position of authorship of mentee in copublications                       \\
\multicolumn{1}{c|}{}                                       & \textit{mentor\_min\_position}                    & min position of authorship of mentor in copublications                       \\
\multicolumn{1}{c|}{}                                       & \textit{mentee\_max\_position}                    & max position of authorship of mentee in copublications                       \\
\multicolumn{1}{c|}{}                                       & \textit{mentor\_max\_position}                    & max position of authorship of mentor in copublications                       \\
\multicolumn{1}{c|}{}                                       & \textit{mentee\_avg\_position}                    & avg position of authorship of mentee in copublications                       \\
\multicolumn{1}{c|}{}                                       & \textit{mentor\_avg\_position}                    & avg position of authorship of mentor in copublications                       \\ \hline
\multicolumn{1}{c|}{\multirow{20}{*}{\textbf{graph}}}       & \textit{coa\_out\_min}                            & coauthor out-edge min weight                                                 \\
\multicolumn{1}{c|}{}                                       & \textit{coa\_in\_min}                             & coauthor in-edge min weight                                                  \\
\multicolumn{1}{c|}{}                                       & \textit{mte\_out\_min}                            & mentee out-edge min weight                                                   \\
\multicolumn{1}{c|}{}                                       & \textit{mte\_in\_min}                             & mentee in-edge min weight                                                    \\
\multicolumn{1}{c|}{}                                       & \textit{coa\_out\_max}                            & coauthor out-edge max weight                                                 \\
\multicolumn{1}{c|}{}                                       & \textit{coa\_in\_max}                             & coauthor in-edge max weight                                                  \\
\multicolumn{1}{c|}{}                                       & \textit{mte\_out\_max}                            & mentee out-edge max weight                                                   \\
\multicolumn{1}{c|}{}                                       & \textit{mte\_in\_max}                             & mentee in-edge max weight                                                    \\
\multicolumn{1}{c|}{}                                       & \textit{coa\_out\_sum}                            & sum of out-edge weights for coauthor                                         \\
\multicolumn{1}{c|}{}                                       & \textit{coa\_in\_sum}                             & sum of in-edge weights for coauthor                                          \\
\multicolumn{1}{c|}{}                                       & \textit{mte\_out\_sum}                            & sum of out-edge weights for mentee                                           \\
\multicolumn{1}{c|}{}                                       & \textit{mte\_in\_sum}                             & sum of in-edge weights for mentee                                            \\
\multicolumn{1}{c|}{}                                       & \textit{mte\_weight\_sum}                         & mte\_out\_sum + mte\_in\_sum                                                 \\
\multicolumn{1}{c|}{}                                       & \textit{coa\_weight\_sum}                         & coa\_out\_sum + coa\_in\_sum                                                 \\
\multicolumn{1}{c|}{}                                       & \textit{mte\_avg\_in}                             & average of in-edge weights for mentee                                        \\
\multicolumn{1}{c|}{}                                       & \textit{mte\_avg\_out}                            & average of out-edge weights for mentee                                       \\
\multicolumn{1}{c|}{}                                       & \textit{coa\_avg\_in}                             & average of in-edge weights for coauthor                                      \\
\multicolumn{1}{c|}{}                                       & \textit{coa\_avg\_out}                            & average of out-edge weights for coauthor                                     \\
\multicolumn{1}{c|}{}                                       & \textit{mte\_ratio\_in\_out}                      & ratio of mte\_in\_sum to mte\_out\_sum                                       \\
\multicolumn{1}{c|}{}                                       & \textit{coa\_ratio\_in\_out}                      & ratio of coa\_in\_sum to coa\_out\_sum                                      
\end{tabular}
\end{small}
\end{table*}

    
    
    
    

\begin{table*}[t]
\caption{Complete results from the fitted \texttt{statsmodels} \cite{seabold2010statsmodels} negative binomial GLM. Note that for field of study, ``unknown'' was used as the reference 0 coefficient category. Note further that before fitting, 99 percentile outlier removal was performed on paper count, citation count and the two sum covariates.}
\label{appendix:glm}
\begin{tabular}{lrrrrrr}
\textbf{Covariate} &     \textbf{Coef.} &  \textbf{Std. Err.} &            \textbf{z} &          \textbf{P>|z|} &   \textbf{[0.025} &    \textbf{0.975]} \\
\midrule
Intercept                                          & -1.634325 &  0.003414 &  -478.698723 &   0.000000e+00 & -1.641016 & -1.627633 \\
Agricultural And Food Sciences           &  0.140752 &  0.003383 &    41.609452 &   0.000000e+00 &  0.134122 &  0.147382 \\
Art                                  & -0.008203 &  0.007007 &    -1.170767 &   2.416926e-01 & -0.021937 &  0.005530 \\
Biology                                 &  0.140790 &  0.003033 &    46.423433 &   0.000000e+00 &  0.134846 &  0.146734 \\
Business                                &  0.089883 &  0.004022 &    22.348532 &  1.247803e-110 &  0.082000 &  0.097765 \\
Chemistry                               &  0.164536 &  0.003305 &    49.790898 &   0.000000e+00 &  0.158059 &  0.171013 \\
Computer Science                     &  0.130829 &  0.003319 &    39.413902 &   0.000000e+00 &  0.124324 &  0.137335 \\
Economics                               &  0.103034 &  0.004238 &    24.310960 &  1.501202e-130 &  0.094728 &  0.111341 \\
Education                              &  0.085535 &  0.004309 &    19.849071 &   1.122578e-87 &  0.077089 &  0.093981 \\
Engineering                              &  0.003292 &  0.003321 &     0.991141 &   3.216166e-01 & -0.003218 &  0.009801 \\
Environmental Science                  &  0.108753 &  0.003325 &    32.708699 &  1.174727e-234 &  0.102236 &  0.115270 \\
Geography                               &  0.150294 &  0.018407 &     8.165211 &   3.208731e-16 &  0.114218 &  0.186371 \\
Geology                                  &  0.129722 &  0.004449 &    29.154761 &  7.270255e-187 &  0.121001 &  0.138442 \\
History                                  &  0.007579 &  0.007482 &     1.013009 &   3.110559e-01 & -0.007085 &  0.022244 \\
Law                                     &  0.054235 &  0.010239 &     5.296658 &   1.179416e-07 &  0.034166 &  0.074303 \\
Linguistics                            &  0.055035 &  0.010163 &     5.415055 &   6.126987e-08 &  0.035115 &  0.074954 \\
Materials Science                       &  0.131333 &  0.003322 &    39.528443 &   0.000000e+00 &  0.124821 &  0.137845 \\
Mathematics                             &  0.172346 &  0.004922 &    35.015816 &  1.292697e-268 &  0.162699 &  0.181993 \\
Medicine                                &  0.116418 &  0.002956 &    39.380783 &   0.000000e+00 &  0.110624 &  0.122212 \\
Philosophy                              &  0.028260 &  0.012606 &     2.241714 &   2.497989e-02 &  0.003552 &  0.052968 \\
Physics                               &  0.109517 &  0.003219 &    34.027028 &  8.877297e-254 &  0.103209 &  0.115825 \\
Political Science                      &  0.096860 &  0.006898 &    14.040819 &   8.769281e-45 &  0.083339 &  0.110381 \\
Psychology                             &  0.159464 &  0.003762 &    42.383743 &   0.000000e+00 &  0.152090 &  0.166838 \\
Sociology                              &  0.102675 &  0.009682 &    10.604561 &   2.837925e-26 &  0.083698 &  0.121651 \\
Paper Count (2nd Quintile) &  0.207493 &  0.001534 &   135.256638 &   0.000000e+00 &  0.204487 &  0.210500 \\
Paper Count (3rd Quintile). &  0.330930 &  0.001603 &   206.492021 &   0.000000e+00 &  0.327788 &  0.334071 \\
Paper Count (4th Quintile). &  0.440313 &  0.001742 &   252.711182 &   0.000000e+00 &  0.436898 &  0.443728 \\
Paper Count (5th Quintile) &  0.696573 &  0.002017 &   345.349712 &   0.000000e+00 &  0.692620 &  0.700526 \\
Citation Count (2nd Quintile) &  1.828905 &  0.001861 &   982.650323 &   0.000000e+00 &  1.825257 &  1.832553 \\
Citation Count (3rd Quintile) &  2.461976 &  0.001868 &  1317.670812 &   0.000000e+00 &  2.458314 &  2.465638 \\
Citation Count (4th Quintile) &  2.917600 &  0.001933 &  1509.490243 &   0.000000e+00 &  2.913812 &  2.921388 \\
Citation Count (5th Quintile) &  3.509647 &  0.002063 &  1700.923846 &   0.000000e+00 &  3.505602 &  3.513691 \\
Menteeship Sum (2nd Quntile) &  0.057742 &  0.001434 &    40.266133 &   0.000000e+00 &  0.054931 &  0.060552 \\
Menteeship Sum (3rd Quntile) &  0.080814 &  0.001419 &    56.963054 &   0.000000e+00 &  0.078034 &  0.083595 \\
Menteeship Sum (4th Quntile) &  0.105018 &  0.001429 &    73.485605 &   0.000000e+00 &  0.102217 &  0.107818 \\
Menteeship Sum (5th Quntile) &  0.138352 &  0.001477 &    93.700988 &   0.000000e+00 &  0.135458 &  0.141246 \\
Menteeship Mean (2nd Quntile) & -0.020535 &  0.001287 &   -15.950493 &   2.826550e-57 & -0.023058 & -0.018012 \\
Menteeship Mean (3rd Quntile) & -0.032761 &  0.001339 &   -24.473922 &  2.800337e-132 & -0.035385 & -0.030138 \\
Menteeship Mean (4th Quntile) & -0.035933 &  0.001395 &   -25.765417 &  2.165674e-146 & -0.038666 & -0.033200 \\
Menteeship Mean (5th Quntile) & -0.023244 &  0.001444 &   -16.100688 &   2.522834e-58 & -0.026073 & -0.020414 \\
Mentorship Sum (2nd Quntile) & -0.005399 &  0.001882 &    -2.868472 &   4.124601e-03 & -0.009088 & -0.001710 \\
Mentorship Sum (3rd Quntile) & -0.012412 &  0.002341 &    -5.301161 &   1.150686e-07 & -0.017001 & -0.007823 \\
Mentorship Sum (4th Quntile) & -0.012612 &  0.002635 &    -4.785979 &   1.701561e-06 & -0.017776 & -0.007447 \\
Mentorship Sum (5th Quntile) &  0.098365 &  0.002937 &    33.497248 &  5.285543e-246 &  0.092609 &  0.104120 \\
Mentorship Mean (2nd Quntile) &  0.032508 &  0.001870 &    17.386650 &   1.041458e-67 &  0.028843 &  0.036172 \\
Mentorship Mean (3rd Quntile) &  0.046286 &  0.002297 &    20.147111 &   2.852842e-90 &  0.041783 &  0.050789 \\
Mentorship Mean (4th Quntile) &  0.048493 &  0.002584 &    18.766996 &   1.406156e-78 &  0.043428 &  0.053557 \\
Mentorship Mean (5th Quntile) &  0.025579 &  0.002768 &     9.240218 &   2.459948e-20 &  0.020153 &  0.031005 \\

\end{tabular}
\end{table*}

\begin{figure*}[hbtp]

    \begin{lstlisting}[language=Python,frame=single]
HYPEROPT_SPACE = { 
    "learning_rate": hp.choice("learning_rate", [0.1, 0.05, 0.01, 0.005, 0.001]),
    "num_leaves": scope.int(2 ** hp.quniform("num_leaves", 2, 7, 1)),
    "colsample_bytree": hp.quniform("colsample_bytree", 0.4, 1, 0.1),
    "subsample": hp.quniform("subsample", 0.4, 1, 0.1),
    "min_child_samples": scope.int(2 ** hp.quniform("min_child_samples", 0, 7, 1)),
    "min_child_weight": 10 ** hp.quniform("min_child_weight", -6, 0, 1),
    "reg_alpha": hp.choice("reg_alpha", [0, 10 ** hp.quniform("reg_alpha_pos", -6, 1, 1)]),
    "reg_lambda": hp.choice("reg_lambda", [0, 10 ** hp.quniform("reg_lambda_pos", -6, 1, 1)]),
    "max_depth": scope.int(hp.choice("max_depth", 
        [-1, 2 ** hp.quniform("max_depth_pos", 1, 4, 1)]
    ))
}
    \end{lstlisting}
    \caption{We optimized the LightGBM models for both stages by running \texttt{hyperopt} for 50 iterations over the search space above.}
    \label{fig:hyperopt}
\end{figure*}

\end{document}